\documentclass[twocolumn,showpacs,aps,floatfix,superscriptaddress]{revtex4}
\usepackage{amsmath,amssymb,graphicx}
\begin{document}
\title{Stratification in the Preferential Attachment Network}
\author{E.~Ben-Naim}
\affiliation{Theoretical Division and Center for Nonlinear
Studies, Los Alamos National Laboratory, Los Alamos, New Mexico
87545 USA}
\author{P.~L.~Krapivsky}
\affiliation{Department of Physics,
Boston University, Boston, Massachusetts 02215 USA}
\begin{abstract}
We study structural properties of trees grown by preferential
attachment. In this mechanism, nodes are added sequentially and
attached to existing nodes at a rate that is strictly proportional to
the degree.  We classify nodes by their depth $n$, defined as the
distance from the root of the tree, and find that the network
is strongly stratified.  Most notably, the distribution
$f_k^{(n)}$ of nodes with degree $k$ at depth $n$ has a power-law
tail, $f_k^{(n)}\sim k^{-\gamma(n)}$. The exponent grows linearly with
depth, $\gamma(n)=2+\frac{n-1}{\langle n-1\rangle}$, where the
brackets denote an average over all nodes.  Therefore, nodes that are
closer to the root are better connected, and moreover, the degree
distribution strongly varies with depth. Similarly, the in-component
size distribution has a power-law tail and the characteristic exponent
grows linearly with depth. Qualitatively, these behaviors extend to a
class of networks that grow by a redirection mechanism.
\end{abstract}
\maketitle

\section{Introduction}

Unlike the ordered crystalline structure of solid-state matter
\cite{ck}, a wide array of natural and man-made networks ranging from
chemical reaction pathways and social groups to the Internet and
airline routes have a strongly heterogeneous structure
\cite{ws,ab,dm}.  The connectivity of an element in such complex
networks varies across multiple scales: most of the nodes have a very
small number of connections, but there is also a small number of
highly connected hubs.

The degree distribution measures the connectivity in the network and
is widely used to characterize the structure of complex networks. This
distribution uniformly samples all nodes in the network. Yet, given
the highly heterogeneous structure of complex networks, it is
plausible that different subsets of nodes have very different
structural characteristics. In this study, we show that the degree
distribution strongly varies in a given network.

We focus on the basic preferential attachment mechanism \cite{has,ba}
which provides a useful model of growing networks \cite{krl,dms}. The
preferential attachment network, where nodes are added sequentially
and attached to existing nodes at a rate proportional to the degree,
demonstrates how a ``rich-gets-richer'' mechanism generates networks
with highly connected nodes and with a broad distribution of
connectivities.

For the preferential attachment network, which has a tree topology, it
is natural \cite{kr01,kcmdsh} to classify nodes by their depth $n$, defined
as the distance from the root. This depth representation divides the
network into layers: the first layer includes nodes with depth $n=1$,
the second layer includes nodes with depth $n=2$, etc.  Our main
result is that the distribution $f_k^{(n)}$ of nodes with degree $k$
at the $n$th layer has an algebraic tail with a depth-dependent
exponent,
\begin{equation}
\label{main}
f_k^{(n)}\sim k^{-\gamma(n)},\qquad
\gamma(n)=2+\frac{n-1}{\langle n-1\rangle}.
\end{equation}
Interestingly, the exponent $\gamma$ grows linearly with depth, and
thus, nodes that are closer to the root tend to be better
connected. The degree distribution \eqref{main} matches the total
degree distribution, $f_k^{\rm total}\sim k^{-3}$, only at the average
depth, $n=\langle n\rangle$. The tail of the degree distribution is
overpopulated with respect to the total distribution below the average
depth and conversely, the tail is underpopulated above the average
depth. Therefore, the network is stratified. In particular, the structure 
of the network changes with the depth because the degree distribution 
is not uniform across the network.

This qualitative behavior extends to other structural features of the
network. In particular, the in-component size distribution that
measures the total number of nodes that emanate from a given node has
a very similar behavior as in any layer the distribution has an
algebraic tail and the corresponding exponent grows linearly with
depth. Furthermore, the tail of the in-component size distribution is
shallow below the average depth but steep above the average depth. We
generalize these results to the broader class of redirection networks,
and find similar network structures.

The rest of this paper is organized as follows. We describe the
preferential attachment network and discuss the distribution of depth
in section II. We analyze the depth dependence of the degree
distribution and the in-component size distribution in sections III
and IV, respectively.  We briefly discuss redirection networks in
section V and conclude in section VI. The correction to the leading
asymptotic behavior in large but finite networks is derived in
Appendix A.

\section{The Preferential Attachment Network}

In the preferential attachment model of network growth, nodes are
added one at a time. Each new node is linked to a target node that is
selected with probability that is strictly proportional to the degree.
This attachment mechanism favors strongly connected nodes over weakly
connected ones. We assume that initially there is a single node, the
root. Since each attachment event adds one node and one link, the
network maintains a tree topology.

\begin{figure}[t]
\includegraphics[width=0.14\textwidth]{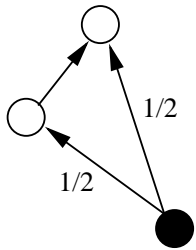}
\caption{Illustration of the redirection process. The new node,
indicated by a bullet, links with equal probabilities to a randomly
selected node or to its parent.}
\label{fig-redirection}
\end{figure}

We use link redirection \cite{kr01} to emulate preferential
attachment. In the redirection process, following the addition of a
new node, an existing node is selected at random. With probability
$\frac{1}{2}$, the new node links to this randomly selected node, and
with equal probability $\frac{1}{2}$, the new node links to the parent
of the selected node, as in figure 1.  A node with degree $k$ has one
outgoing link and $k-1$ incoming links, and the total probability
$P_k$ of attachment to such a node includes two contributions: the
probability of a direct link is $\frac{1}{2N}$ where $N$ is the total
number of nodes while the probability of a redirected link is
$\frac{k-1}{2N}$.  Thus, the total attachment probability
$P_k=\frac{k}{2N}$ is strictly proportional to the degree, and this
link redirection process is equivalent to preferential attachment.
Redirection does not explicitly involve the degree of a node, and is
therefore convenient for both theoretical analysis and numerical
simulation \cite{kr01,kr02}.

Let us label the nodes in the network by their depth $n$, defined as
the distance from the root. With this definition, nodes
are grouped by layers: the first layer consists of nodes with $n=1$,
the second layer consists of nodes with $n=2$, etc, as illustrated in
figure 2.  As a preliminary step, we evaluate $M_n(N)$, the expected
number of nodes at the $n$th layer in a network with $N$ nodes. This
quantity obeys the difference equation
\begin{equation}
\label{Mn-eq-diff}
M_n(N+1)-M_n(N)=\frac{M_{n-1}(N)+M_n(N)}{2N}.
\end{equation}
The boundary condition is $M_0=1$ because there is a single root. The
first gain term on the right hand side accounts for direct links and
the second term for redirected links. Also, the right hand side is
inversely proportional to the total number of nodes. This difference
equation can be converted into a differential equation when the
network is large, $N\gg 1$,
\begin{equation}
\label{Mn-eq}
\frac{dM_n}{dN}=\frac{M_{n-1}+M_n}{2N}.
\end{equation}
Henceforth, the $N$-dependence is implicit.  The number of nodes at
the first layer, \hbox{$M_1\sim N^{1/2}$}, follows from
\hbox{$dM_1/dN=M_1/(2N)$} \cite{kr01}. In general, the transformation
\hbox{$M_n=M_1 m_n$} reduces equation \eqref{Mn-eq} to
\hbox{$dm_n/dN=m_{n-1}/(2N)$} with the boundary condition
$m_1=1$. Solving this latter equation recursively yields
\hbox{$m_2=\frac{1}{2}\ln N$},
\hbox{$m_3=\frac{1}{2!}\left(\frac{1}{2}\ln N\right)^2$}, and in
general, \hbox{$m_n=\frac{1}{(n-1)!}\left(\frac{1}{2}\ln
N\right)^{n-1}$}. Therefore, the distribution of depth is
\begin{equation}
\label{Mn}
M_n\sim N^{1/2}\frac{\left(\frac{1}{2}\ln N\right)^{n-1}}{(n-1)!}.
\end{equation}
Accordingly, the distribution of the variable $n-1$ is Poissonian and
is fully characterized by the average that grows logarithmically
with the total number of nodes,
\begin{equation}
\label{nav}
\langle n-1 \rangle \simeq\frac{1}{2}\ln N.
\end{equation}
Furthermore, the variance, $\sigma^2=\langle (n-1)^2\rangle -\langle
n-1\rangle^2$, equals the average, $\sigma^2\simeq\frac{1}{2}\ln N$.

\begin{figure}[t]
\includegraphics[width=0.35\textwidth]{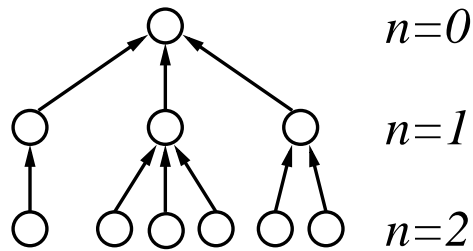}
\caption{Illustration of the layer structure. Indicated in the figure
are the  first three layers.}
\label{fig-layers}
\end{figure}

In principle, the Poisson depth distribution \eqref{Mn} is Gaussian in
the vicinity of the average depth. Specifically, the the variable
$\xi=[(n-1)-\langle n-1\rangle]/\sigma$ obeys Gaussian statistics.  In
practice, since the depth and the variance both grow logarithmically
with the network size, statistics of the depth are not fully captured
by the Gaussian distribution even in very large networks.

We measure the depth in units of the average using the normalized
depth
\begin{equation}
\label{y-def}
y=\frac{n-1}{\langle n-1\rangle}.
\end{equation}
The average number of nodes with normalized depth $y$ grows as a
power-law with the total number of nodes,
\begin{equation}
\label{Mx}
M_n\sim N^\nu, \qquad \nu=\frac{1+y-y\ln y}{2}.
\end{equation}
This result is obtained by substituting $ n-1= \frac{y}{2}\ln N$ into
\eqref{Mn} and then evaluating the quantity $\ln M_n\simeq \nu\ln N$
using the Stirling formula, $\ln n!\sim n\ln n-n$. As expected, the
total number of nodes at the average depth is proportional to the
system size, $\nu(1)=1$, and the total number of nodes at the first
layer is consistent with \eqref{Mn}, $\nu(0)=\frac{1}{2}$. In general,
the exponent $\nu$ that characterizes the population of nodes at a
given depth is continuously varying, and in particular, this exponent
vanishes at a nontrivial maximal depth, $y_{\rm max}=3.59112$,
specified as the root of the equation $1+y_{\rm max}=y_{\rm max}\ln
y_{\rm max}$. Therefore, there are no nodes with depth larger than
this maximal depth, $0\leq y\leq y_{\rm max}$.

\section{The Degree Distribution}

We now discuss the unnormalized degree distribution, $F_k^{(n)}$,
defined as the average number of nodes with depth $n$ and degree
$k$. We begin with the first layer, $n=1$, where the quantity
$F_k^{(1)}$ obeys the rate equation
\begin{equation}
\label{Fk1-eq}
\frac{dF_k^{(1)}}{dN} = \frac{(k-1)F_{k-1}^{(1)} - kF_k^{(1)}}{2N} +
\frac{M_0+M_1}{2N}\,\delta_{k,1}.
\end{equation}
Throughout this study, we assume that the network is large and treat
$N$ as a continuous variable. In other words, we use rate equations as
in \eqref{Mn-eq} rather than difference equations as in
\eqref{Mn-eq-diff}. The first two gain terms on the right hand side
account for augmentation in the degree of an existing node due to
attachment, and the corresponding rate of attachment to nodes with
degree $k$ equals the attachment probability $P_k=\frac{k}{2N}$. The
last two gain terms, accounting respectively for redirected links and
direct links, correspond to the new nodes.

We now introduce the degree distribution $f_k^{(1)}$, the fraction
 first layer nodes with degree $k$, defined by
\begin{equation}
\label{fk1-def}
F_k^{(1)} = M_1 f_k^{(1)}.
\end{equation}
This distribution is normalized, $\sum_k f_k^{(1)}=1$.  With the
definition \eqref{fk1-def}, the evolution equation \eqref{Fk1-eq}
becomes
\begin{eqnarray*}
\frac{M_0+M_1}{2N}f_k^{(1)}=
M_1\frac{(k-1)f_{k-1}^{(1)} - kf_k^{(1)}}{2N} +
\frac{M_0+M_1}{2N}\,\delta_{k,1}.
\end{eqnarray*}
In writing this equation, we kept the rate of change of the total
number of nodes in the first layer $dM_1/dN$ given by \eqref{Mn-eq},
but neglected the sub-dominant $N$-dependence of the normalized degree
distribution $df_k^{(1)}/dN$. The number of nodes at the zeroth layer,
$M_0=1$, is negligible compared with the number of nodes at the first
layer, $M_0\ll M_1$, and as a result, the degree distribution
$f_k^{(1)}$ obeys the simple recursion equation
\begin{equation}
\label{fk1-eq}
(k+1)f_k^{(1)} = (k-1)f_{k-1}^{(1)} + \delta_{k,1}.
\end{equation}
Hence, the degree distribution is $f_k^{(1)}=\frac{1}{1\cdot 2},
\frac{1}{2\cdot 3}, \frac{1}{3\cdot 4}$ for $k=1,2,3$, and in general,
this quantity is remarkably simple,
\begin{equation}
\label{fk1}
f_k^{(1)} = \frac{1}{k(k+1)}.
\end{equation}
This distribution is properly normalized, \hbox{$\sum_k f_k^{(1)}=1$}.
Interestingly, the tail of the {\em normalized} degree distribution,
\hbox{$f_k^{(1)}\sim k^{-2}$} is overpopulated with respect to that of
the total degree distribution, \hbox{$f_k^{\rm total}\sim k^{-3}$}
\cite{ba,krl,dms}.  Nevertheless, the expected number of nodes with
degree $k$ at the first layer, \hbox{$F_k^{(1)}\sim N^{1/2}k^{-2}$},
remains smaller than the total number of nodes with degree $k$,
\hbox{$F_k^{\rm tot}\sim Nk^{-3}$}, because the maximal degree in the
network is bounded, \hbox{$k\ll N^{1/2}$}.  The maximal degree in the
network, $k_{\rm max}\sim N^{1/2}$, is estimated by equating the
cumulative degree distribution \hbox{$Q_k=\sum_{j\geq k} F^{\rm tot}_j\sim
Nk^{-2}$ to one, $Q_k\sim 1$}. Equation \eqref{fk1} clearly shows that
nodes at the first layer tend to be much more connected compared with
the rest of the network.

\begin{figure}[t]
\includegraphics[width=0.4\textwidth]{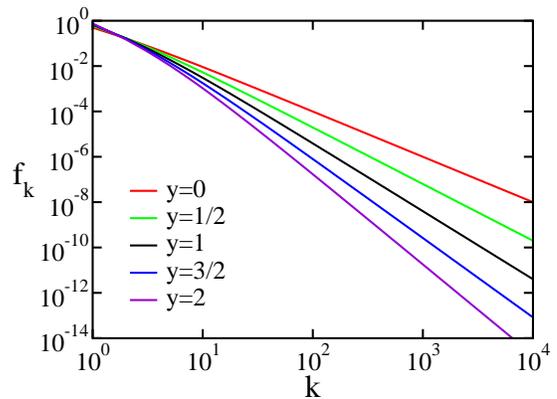}
\caption{The degree distribution at different depths.  Shown is the
closed form expression \eqref{fkn} for five, equally-spaced, values of
the normalized depth $y$ defined in \eqref{y-def}.}
\label{fig-lattice}
\end{figure}

The equation governing $F_k^{(n)}$, the average number of
nodes with degree $k$ at the $n$th layer, is given by a
straightforward generalization of \eqref{Fk1-eq},
\begin{equation}
\label{Fkn-eq}
\frac{d F_k^{(n)}}{dN} = \frac{(k-1)F_{k-1}^{(n)} - kF_k^{(n)}}{2N}
+ \frac{M_{n-1}+M_n}{2N}\,\delta_{k,1}.
\end{equation}
Following the steps leading to \eqref{fk1-eq}, the degree
distribution, $f_k^{(n)}$, defined as in \eqref{fk1-def}, $F_k^{(n)} =
M_n f_k^{(n)}$, obeys the recursion equation
\begin{equation}
\label{fkn-eq}
(k+1+y)f_k^{(n)} = (k-1)f_{k-1}^{(n)} + (1+y)\delta_{k,1}
\end{equation}
with the shorthand notation $y=M_{n-1}/M_n$.  The
parameter $y$ is equivalent to the normalized depth defined in \eqref{y-def}.
The degree distribution follows immediately from
the recursion equation \eqref{fkn-eq}. First, the fraction of leafs is
\hbox{$f_1^{(n)}=(1+y)(2+y)$}. Second, using the ratio
\hbox{$f_k^{(n)}/f_{k-1}^{(n)}=(k-1)(k+1+y)$} and the following
property of the Gamma function, \hbox{$\Gamma(x+1)/\Gamma(x)=x$}, we
express the degree distribution in a closed form in terms of the Gamma
function,
\begin{equation}
\label{fkn}
f_k^{(n)}=(1+y)
\frac{\Gamma(2+y)\Gamma(k)}{\Gamma(k+2+y)}.
\end{equation}
Therefore, there is a family of degree distributions, parameterized by
the normalized depth $y$.  From the well-known asymptotic behavior of
the ratio of Gamma functions, $\lim_{x\to\infty}[x^a\Gamma(x)/\Gamma(x+a)]=1$,
we find that the degree distribution has a power-law tail,
\begin{equation}
\label{fkn-tail}
f_k^{(n)}\simeq A\,k^{-\gamma(n)}
\end{equation}
for $k\gg 1$. The characteristic exponent $\gamma=2+y$ grows linearly
with depth
\begin{equation}
\label{gamma}
\gamma(n)=2+\frac{n-1}{\langle n-1\rangle},
\end{equation}
and the prefactor is $A=(1+y)\,\Gamma(2+y)$. The power-law behavior
\eqref{fkn-tail} strictly holds only for infinitely large networks. The
appendix describes the correction to this leading asymptotic behavior
in large but finite networks.

As shown in figure 3, the degree distribution is not uniform across
the network and the characteristic exponent $\gamma(n)$ increases
linearly with depth $n$. Therefore, nodes that are closer to the root
tend to have a larger number of connections. The tail of the degree
distribution is overpopulated with respect to the total degree
distribution below the average depth as $\gamma<3$ for $n<\langle
n\rangle$.  Similarly, the tail is underpopulated with respect to the
total degree distribution above the average depth, $\gamma>3$ for
$n>\langle n\rangle$.  The degree distribution at the average depth
equals the total degree distribution \hbox{$f_k^{(\langle
n\rangle)}=f_k^{\rm total}=4/[k(k+1)(k+2)]$}
\label{ba,krl,dms} because the depth distribution \eqref{Mn}
gradually narrows around the average depth \eqref{nav} as the network
grows. Nevertheless, there is still an appreciable number of nodes at
depths other than the average, as indicated by the power-law growth in
\eqref{Mx}. Finally, we note that since the depth is bounded by the
maximum value $y_{\rm max}=3.59112$, the characteristic
exponent has a nontrivial upper bound, $2\leq \gamma\leq 5.59112$.

\section{The in-component size distribution}

Each node is in itself a root of the sub-tree that emanates from it. This
sub-tree is termed the in-component of the node and we denote by $s$
the size of the in-component. We note that the size of the in-component
is at least as large as the degree of the node: $s\geq k$.  The
minimal size $s=1$ corresponds to dangling nodes without descendants.
Let $G_s^{(n)}$ be the average number of $n$th layer nodes with
in-components of size $s$. This quantity satisfies the rate equation
\begin{equation*}
\frac{dG_s^{(n)}}{dN} =
\frac{\left(s-\frac{3}{2}\right)G_{s-1}^{(n)}-
\left(s-\frac{1}{2}\right)G_s^{(n)}}{N}+
\frac{M_{n-1}+M_n}{2N}\delta_{s,1}.
\end{equation*}
With probability $\frac{s}{N}$ a node in an in-component of size $s$
is selected at random following the addition of a new node, and all
such events, with the exception of redirection away from the root of
the sub-tree, result in attachment of an additional node to the
in-component. Since redirection away from the root of the sub-tree
occurs with probability $\frac{1}{2N}$, the probability of attaching
an additional node to the in-component equals
$\frac{s-1/2}{N}$. Hence, the first two terms on the right hand
side. The last two terms are as in \eqref{Fkn-eq}.

The normalized distribution $g_s^{(n)}$ of nodes with in-degree $s$
and depth $n$, defined by $G_s^{(n)} = M_n\,g_s^{(n)}$, obeys the
recursion equation
\begin{equation}
\label{gsn-eq}
\left(s +\frac{y}{2}\right) g_s^{(n)} =
\left(s-\frac{3}{2}\right)g_{s-1}^{(n)}+\frac{1+y}{2}\delta_{s,1}~.
\end{equation}
This equation is very similar in structure to the equation
\eqref{fkn-eq} governing the degree distribution. Dangling nodes have
$s=k=1$ and therefore,
\hbox{$g_1^{(n)}=f_1^{(n)}=(1+y)(2+y)$}. Using the ratio
\hbox{$g_s^{(n)}/g_{s-1}^{(n)}=(y-3/2)/(s+y/2)$}, we obtain the
in-component size distribution in closed form
\begin{equation}
\label{gsn}
g_s^{(n)}=
\frac{1+y}{2+y}
\frac{\Gamma\left(2+\frac{y}{2}\right)}{\Gamma\left(\frac{1}{2}\right)}
\frac{\Gamma\left(s-\frac{1}{2}\right)}{\Gamma\left(s+1+\frac{y}{2}\right)}~.
\end{equation}
Again, there is a family of distributions that is governed by $y$.
Therefore, the in-component size distribution has a power-law tail
\begin{equation}
\label{gsn-tail}
G_s^{(n)}\simeq B\,s^{-\beta(n)}
\end{equation}
and the characteristic exponent varies continuously with depth,
$\beta=(3+y)/2$ or
\begin{equation}
\label{beta}
\beta(n)=\frac{3}{2}+\frac{1}{2}\frac{n-1}{\langle n-1\rangle}.
\end{equation}
The prefactor in \eqref{gsn-tail} is $B=(1+y)\Gamma(1+y/2)/\sqrt{4\pi}$.

\begin{figure}[t]
\includegraphics[width=0.4\textwidth]{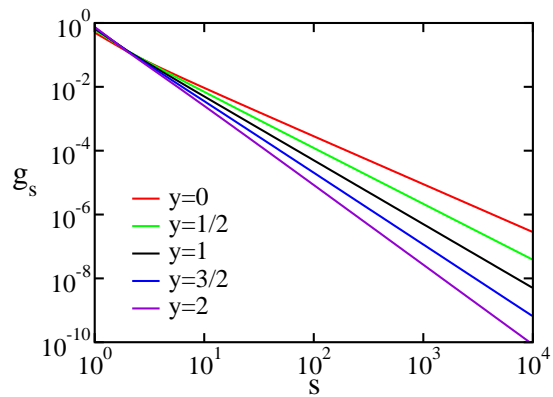}
\caption{The in-component size distribution at different depths. Shown is
the closed form expression \eqref{gsn} for $y=0,1/2,1,3/2,2$.}
\end{figure}

The in-component size distribution is very similar to the degree
distribution (Figure 4). Nodes that are closer to the root tend to
have larger in-components. The tail of the in-component size
distribution is overpopulated below the average depth, $\beta<2$ when
$n<\langle n\rangle$, while the tail is underpopulated above the
average depth, $\beta>2$ for $n>\langle n\rangle$. The in-component
size distribution matches the total distribution at the average depth,
\hbox{$G_s^{(\langle n\rangle)}=G_s^{{\rm total}}=2\,[4s^2-1]^{-1}$}.
Moreover, the in-component size distribution is steepest, $\beta_{\rm
max}=4.29556$, at the maximal depth $y_{\rm max}=5.59112$.

For completeness, we also quote the in-component size distribution in the
first layer,
\begin{equation}
\label{gs1}
g_s^{(1)}=\frac{1}{\sqrt{4\pi}}
\frac{\Gamma\left(s-\frac{1}{2}\right)}{\Gamma(s+1)}~,
\end{equation}
and the corresponding tail,
$g_s^{(1)}\sim s^{-3/2}$.

\section{Redirection networks}

We briefly discuss the broader class of redirection networks
\cite{kr01}. This family of networks is also grown by sequential
addition of nodes. Subsequent to the addition of a new node, one
existing node is selected at random. With the redirection probability
$r$, the new node attaches to the parent of the randomly selected node
while with the complementary probability $1-r$, the new nodes attaches
to the randomly selected node itself.  Redirection networks are
therefore parameterized by the redirection probability $r$.  The
probability $P_k$ of attachment to a node of degree $k$ varies
linearly with the degree, $P_k=(k-1)r+(1-r)$ or equivalently,
\begin{equation}
\label{Pk}
P_k=\frac{kr+1-2r}{N}.
\end{equation}
As mentioned above, the special case $r=1/2$ yields the preferential
attachment network. The limiting case \hbox{$r=0$} corresponds to the
random recursive tree where the attachment probability is uniform
\cite{jmr,bp,ld,pn,hmm,dek,fsh}.

The depth distribution obeys the following generalization of \eqref{Mn-eq}
\begin{equation}
\label{Mn-eq-r}
\frac{dM_n}{dN}=\frac{(1-r)M_{n-1}+rM_n}{N}.
\end{equation}
Here, the first gain term accounts for direct links and the second
term, for redirected links.  The depth distribution is always
Poissonian
\begin{equation}
\label{Mn-r}
M_n\sim N^{r}\frac{\left[(1-r)\ln N\right]^{n-1}}{(n-1)!}
\end{equation}
and fully characterized by the average 
\hbox{$\langle n-1\rangle\simeq(1-r)\ln N$}.

The degree distribution is obtained by replacing the attachment
probability $P_k=\frac{k}{2N}$ in \eqref{Fkn-eq} with \eqref{Pk}. We
skip the straightforward derivation of the degree distribution and
merely quote the final result,
\begin{equation}
f_k^{(n)}=
\frac{r+y-ry}{1+y-ry}
\frac{\Gamma\left(\frac{1+r+y-ry}{r}\right)}
{\Gamma\left(\frac{1-r}{r}\right)}
\frac{\Gamma\left(k+\frac{1-2r}{r}\right)}
{\Gamma\left(k+\frac{1+y-ry}{r}\right)}.
\end{equation}
The parameter $y$ is the same normalized depth of \eqref{y-def} with
the appropriate average $\langle n-1\rangle\simeq(1-r)\ln N$. Therefore,
the degree distribution decays algebraically as in \eqref{fkn-tail}
with the exponent $\gamma=2+[(1-r)y]/r$ or equivalently
\begin{equation}
\gamma(n)=2+\frac{1-r}{r}\frac{n-1}{\langle n-1\rangle}.
\end{equation}

The in-component size distribution
\begin{equation}
g_s^{(n)}=
\frac{r+y-ry}{1+y-ry}
\frac{\Gamma(2+y-ry)}{\Gamma(1-r)}
\frac{\Gamma(s-r)}{\Gamma(s+1+y-ry)}
\end{equation}
is obtained by replacing the attachment probability $\frac{s-1/2}{N}$
in \eqref{gsn-eq} with $\frac{s-r}{N}$.  As expected, this
distribution decays algebraically as in \eqref{gsn-tail} with the
exponent
\begin{equation}
\beta(n)=1+r+(1-r)\frac{n-1}{\langle n-1\rangle}.
\end{equation}

There is no substantive change in the behaviors of the degree
distribution and the in-component size distribution. These
distributions are characterized by power-law tails that become steeper
with increasing depth, so that the further from the root a node is,
the less likely that the node is highly connected. Generally, the
degree distribution is non-uniform throughout the network.

\section{conclusions}

In conclusion, we studied how the degree distribution depends on
depth, defined as the distance from the root, in the preferential
attachment network and found that the network is strongly stratified.
There is a family of degree distributions that is parameterized by the
depth and the total degree distribution is a special case that
corresponds to the behavior at the average depth.  The degree
distribution has a shallow power-law tail below the average depth and
a steep tail above the average depth as the characteristic exponent
grows linearly with depth. Interestingly, this exponent has a
non-trivial upper bound.  The in-component size distribution exhibits
very similar qualitative behaviors.

The structure of the network changes considerably with the depth as
nodes that are closer to the root tend to have a larger number of
connections. Such stratification is empirically observed in complex
networks including the internet \cite{chkss}, and can be intuitively
understood as follows.  There are strong correlations between the
depth of a node and its age because young nodes must be less
connected.  Since younger nodes are also further from the root, there
are correlations between the degree and the depth. A natural
complementary classification of the nodes is by their age, and we
anticipate that similarly, the degree distribution exhibits age
dependence.

The above results are not limited to trees. This is seen from a
generalization of the preferential attachment process where new nodes
attach to two nodes: one target node that is selected with probability
proportional to its degree and one, randomly selected, parent of the
target node. Thus, each new node adds a three-node cycle, and the
network has at least as many cycles as there are nodes. Yet, the
governing equations do not change and the behavior of the degree
distribution and in-component size distribution are basically, the
same \cite{bk}.

One issue, open to further investigations, is the behavior in finite
systems. The analysis in the appendix relies on on the continuum
approximation. Yet, this approximation is asymptotically exact for
degrees that are much smaller than the maximal one.  A discrete
approach with difference equations, rather than differential ones, is
necessary \cite{kk,dms2,kr02-a} to determine the behavior for the
degrees that are of the order of the maximal degree.

\acknowledgments We thank Sergei Dorogovtsev for collaboration on the
in-component distribution in the first layer, Eq.~\eqref{gs1}.  We are
grateful for financial support from DOE grant DE-AC52-06NA25396 and
NSF grant CCF-0829541.

\appendix

\section{Finite Networks}

Throughout this investigation, we implicitly considered the leading
asymptotic behavior for large networks. In particular, the
sub-dominant dependence of the normalized degree distribution on the
size of the network was ignored in \eqref{fkn-eq}. However, the
logarithmic dependencies of the average depth and the variance suggest
that extremely large networks may be needed to clearly observe the
large-$N$ asymptotics \eqref{fkn-tail}.

To investigate how the degree distribution depends on the total number
of nodes $N$, we rewrite the governing equation \eqref{Fkn-eq} in
continuum form for large $k$,
\begin{equation}
\label{F-eq}
\frac{\partial F}{\partial N}+
\frac{1}{2N}\frac{\partial }{\partial k}\left(kF\right)=0.
\end{equation}
In deriving this equation, we omitted the superscript and the
subscript, $F\equiv F_k^{(n)}$.  We now include the term $\partial
f/\partial N$ describing how the degree distribution changes with
system size in the evolution equation for the normalized distribution,
defined by $F=Mf$,
\begin{equation}
\label{f-eq}
2N\frac{\partial f}{\partial N}+k\frac{\partial f}{\partial k}+(2+y)f=0.
\end{equation}
The normalized depth \eqref{y-def} is $y=2(n-1)/\ln N$ and therefore,
\hbox{$dy/dN=-y/\left(N\ln N\right)$}. We now assume that the
power-law tail \eqref{fkn-tail} is modified by a correction,
\hbox{$f(k,N)\simeq A\,k^{-(2+y)}u(k,N)$}. From \eqref{f-eq}, the
correction function $u$ satisfies
\begin{equation}
\frac{\partial u}{\partial k}+2y\frac{\ln k}{k\ln N}u=0
\end{equation}
Therefore, the correction function is log-normal,
\hbox{$u=\exp[-2y(\ln k)^2/\ln N]$}, and the normalized distribution
has the following tail
\begin{equation}
\label{f-corrected}
f_k^{(n)}\simeq A\,k^{-2-y}\,\exp\left[-2y\frac{(\ln k)^2}{\ln N}\right].
\end{equation}
Indeed, the correction is irrelevant for infinitely large networks:
$u\to 1$ when $N\to\infty$. For large but finite systems, there are
two consequences. First, at moderate degrees, the correction is
sub-dominant and the tail is power-law, although the exponent may
appear slightly larger than the asymptotic value \eqref{gamma}.
Second, at large degrees, the correction is dominant and the tail is
log-normal.

\end{document}